\documentclass[12pt]{iopart} \usepackage{graphicx}
\usepackage{epstopdf} 
\usepackage{bm} \usepackage{array}
\usepackage[normalem]{ulem}         

\newcommand{\iotabar}{\mbox{$\iota\!\!$-}}

\usepackage{iopams}  

\begin{document}

\title[Overdense microwave heating in CNT]
      {Overdense microwave plasma heating in the CNT stellarator} 

\author{K.C.~Hammond$^1$, R.R.~Diaz-Pacheco$^1$, A.~K\"{o}hn$^2$,
        F.A.~Volpe$^1$, Y. Wei$^1$}

\address{$^1$ Department of Applied Physics and Applied Mathematics, 
             Columbia University, New York, NY 10027, USA}
\address{$^2$ Max Planck Institute for Plasma Physics, 85748 Garching, Germany}

\ead{fvolpe@columbia.edu}


\date{\today}

\begin{abstract}
Overdense plasmas have been attained with 2.45 GHz microwave heating in the 
low-field, low-aspect-ratio CNT stellarator.
Densities higher than four times the ordinary (O) mode cutoff 
density were measured with 8 kW of power injected in the O-mode and,
alternatively, with 6.5 kW in the extraordinary (X) mode. 
The temperature profiles peak at the plasma edge. This was ascribed to 
collisional damping of the X-mode at the upper hybrid resonant layer. 
The X-mode reaches that location by tunneling, mode-conversions or after
polarization-scrambling reflections off the wall and in-vessel coils,
regardless of the initial
launch being in O- or X-mode. This interpretation was confirmed by full-wave
numerical simulations. Also, as the CNT plasma is not completely ionized at
these low microwave power levels, electron density was shown to increase with
power. A dependence on magnetic field strength was also observed (for O-mode 
launch) and discussed.
\end{abstract}

\section{Introduction}
\label{sect:intro}

The need for high triple product and desire for high plasma $\beta$ in fusion 
experiments translate into a simultaneous requirement for high densities 
and temperatures. In devices with relatively low magnetic fields, this 
implies that the ratio $\omega_\mathrm{pe}/\omega_\mathrm{ce}$ of the electron
plasma frequency to the electron cyclotron frequency might exceed unity. In 
this case, the plasma may not admit the propagation of electromagnetic waves in
the electron cyclotron frequency range. Thus, if electron cyclotron heating
(ECH)\cite{prater2004} is desired, overdense heating techniques become 
necessary. 

One common overdense heating mechanism involves mode-conversion of 
electromagnetic to electron Bernstein waves (EBWs) inside the plasma, because
EBW propagation does not suffer from upper density limits.\cite{laqua2007}
Three techniques are commonly
employed in tokamaks, spherical tokamaks, stellarators and
reversed field pinches.

In the SX-B scheme, implemented for example on
WT-3\cite{maekawa2001}, COMPASS-D \cite{shevchenko2002} and
LHD,\cite{yoshimura2013} a wave 
launched from the high-field side in the slow extraordinary (SX) mode 
converts to an EBW upon reaching the upper hybrid resonance (UHR).

In the FX-B scheme, a fast X-mode launched from the low-field side impinges on
its cutoff and mode-converts to an EBW at the UHR after ``tunneling'' through a 
narrow region of evanescence. This scheme has been 
implemented for example on CHS \cite{ikeda2008} and MST.\cite{seltzman2016}

In the O-X-B scheme,\cite{preinhaelter1973} a wave is launched from the 
low-field side in the ordinary (O) mode in a special direction, such that it
reaches the cutoff layer with an optimal value of refractive index
parallel to the magnetic field, $N_\parallel=N_{\parallel,opt}$, and with $N_y$=0,
where $y$ denotes the direction orthogonal to the local magnetic field and
density gradient. 
This special incidence facilitates conversion to the SX mode,
which in turn converts to an EBW at the UHR.
This has been realized for example in W7-AS\cite{laqua2003}
and TCV\cite{mueck2007}, and is planned
for Heliotron J\cite{nagasaki2016}.

Overdense heating has also been achieved in small low-field devices such as 
TJ-K\cite{koehn2010} and WEGA.\cite{podoba2007}
In these experiments, the 
vacuum wavelength $\lambda_0$ associated with $\omega_\mathrm{ce}$ is on the
order of the plasma minor radius $a$. While propagation and damping 
are more difficult to analyze in this regime, full-wave modeling 
verified successful O-X mode conversion in 
WEGA,\cite{podoba2007} and confirmed collisional heating at the UHR as the 
dominant mechanism in TJ-K.\cite{koehn2010}

The low field, low aspect ratio CNT stellarator\cite{pedersen2004} can be 
viewed as a further extension of this long-wavelength regime.
CNT was originally dedicated to non-neutral and quasi-neutral plasma 
research,\cite{kremer2006,sarasola2012} but has recently been re-purposed to
investigate fusion-relevant problems such as error
field analysis,\cite{hammond2016} high-$\beta$ stability,\cite{hammond_pop2017} 
and inversion of stellarator images.\cite{hammond_rsi2016} As part of this, it
was equipped with microwave heating at 2.45 GHz. The corresponding vacuum
wavelength, $\lambda_0=12.2$ cm, is comparable with the plasma minor radius 
$a\simeq13$ cm and about one-third of the major radius $R\simeq30$ cm. 
On this scale, the broad launched 
microwave beam strikes a broad region of the plasma at a range of incident 
angles and polarizations. This variation has been  limited to some extent in
CNT by placing a waveguide very close to the plasma edge. However, no focusing 
element was used in this initial study. Also, the launch system was not yet
optimized for any particular overdense heating mechanism. 

In spite of this, plasmas were attained in CNT that were overdense to 
O-mode propagation by factors of more than 4, as shown in the present paper. 
Sec.~\ref{sect:10kW_heating_diagnostics} describes the heating and diagnostic 
systems employed for this work. Profiles of 
density and temperature confirmed
overdense heating for both O- and X-mode launch. These profiles 
are presented in Sec.~\ref{sect:profiles} alongside a
study of their dependence on power and magnetic field. 
Sec.~\ref{sect:fullWave} describes the full-wave numerical method used to 
interpret the experimental results. Interpretations are discussed
in Sec.~\ref{sect:10kW_discussion}.

\section{Experimental setup}
\label{sect:10kW_heating_diagnostics}

\subsection{Heating system}
\label{subsect:heatingSystem}

\begin{figure} \begin{center}
    \includegraphics[width=0.47\textwidth]{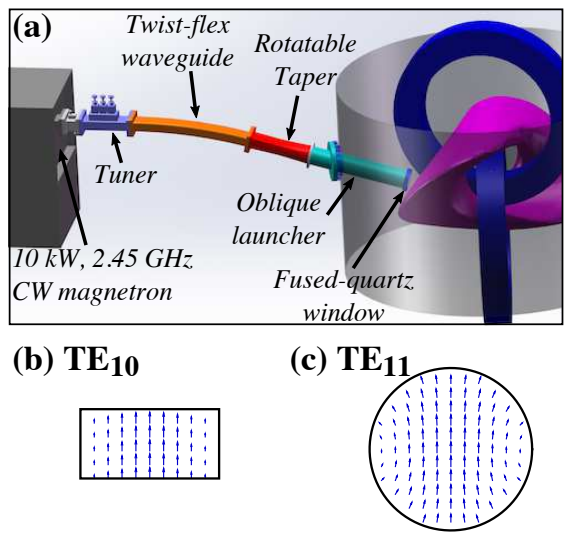}
    \caption{(a) Schematic of the microwave heating system as described in the 
             text.
             (b)-(c) Electric field vectors for the fundamental rectangular
                 waveguide mode and $\mathrm{TE}_{11}$ circular
                 waveguide mode. }
    \label{fig:waveguide_schematic}
\end{center} \end{figure}

The microwave heating system is shown schematically in 
Fig.~\ref{fig:waveguide_schematic}. The microwave source is a 10 kW, 2.45 GHz 
magnetron manufactured by Muegge. At full power, its output is CW.
Reduced power (in a time-averaged sense) is obtained 
by pulsed operation at roughly 5 kHz: the
lower the duty-cycle, the lower the averaged output power. 
The magnetron launcher excites the $\mathrm{TE}_{10}$ 
mode (Fig.~\ref{fig:waveguide_schematic}b) in a rectangular waveguide. 

A three-stub tuner controls the percentage of power coupled to the plasma, and
the reflected power is absorbed in a water-cooled isolator upstream of the tuner
(not shown). The power injected in the vessel is determined by the difference 
of the forward and reflected power as measured by a pair of Schottky diodes
fixed to a dual directional coupler. This net injected power is an upper bound
for the power actually deposited in the plasma. The rest is dissipated on the
resistive walls or leaks out of the vessel through few unshielded ports, after 
multiple reflections off the walls. 

Downstream of the tuner is a twistable, flexible rectangular 
waveguide followed by a rectangular-to-circular taper that can rotate freely 
on the flange on its circular side.
The polarization is linear in the $\mathrm{TE}_{10}$ 
mode in the rectangular waveguide (Fig.~\ref{fig:waveguide_schematic}b). 
This is tapered into the $\mathrm{TE}_{11}$ 
mode in the circular waveguide (Fig.~\ref{fig:waveguide_schematic}c),
which is also linearly polarized, to some approximation. 
By choosing the correct orientation of the twistable, flexible waveguide and of
the rotatable taper, one can rotate the wave electric field 
relative to the magnetic field.
This allows selecting the O- or X-mode as the {\em dominant} 
polarization injected in the plasma. The nominal O and X polarizations
referred to below are not pure because,
due to oblique injection, the actual O and X eigenmodes are 
elliptically polarized.
This can be quantified by the complex ratio $E_{x'}/E_{y'}$ of the wave 
electric field components orthogonal to the direction of propagation. 
From the Appleton-Hartree dispersion, we calculate 
$iE_{x'}/E_{y'}=-2.5-1.2i$ on the beam axis, for O-mode propagation
and magnetic field $|\mathbf{B}| = 87.5$ mT, discussed below. 
Different values apply elsewhere on the wavefront, due to beam divergence
and magnetic field curvature. Nonetheless, the value provided gives
an order of magnitude of the lack of modal purity. 

Following the taper is the launch antenna:
a section of circular waveguide that leads into the vacuum vessel up to near 
the plasma edge. The launch antenna is effectively a re-entrant port, held at
atmospheric pressure to avoid unwanted breakdown and plasma formation
within the waveguide, at locations where the wave-frequency equals
high-order EC harmonics. Such plasma would 
partly absorb or fully reflect the high-power microwaves in the waveguide,
before they reach the stellarator plasma. Both 
effects are undesired. A quartz window at the end of the launch antenna 
functions as the vacuum break.

\begin{figure} \begin{center}
    \includegraphics[width=0.47\textwidth]{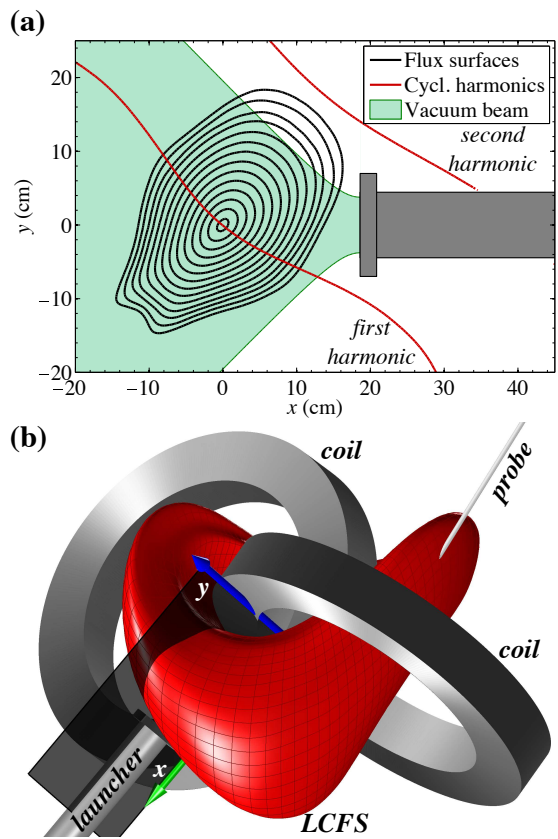}
    \caption{(a) Launch antenna (gray), X-mode Gaussian beam (green), cross-section of 
             flux surfaces (black), and electron cyclotron resonances 
             for 2.45 GHz (red). The beam boundary corresponds to field 
             intensities of $1/e^2$ relative to the values on the beam axis.
             Note that, in this plane, the O-mode beam would be slightly
             wider. The $x$
             axis is coaxial with the waveguide and has its origin in the
             magnetic axis.
             (b) Three-dimensional schematic of
             last closed flux surface (LCFS), launch antenna, 
             Langmuir probe, and in-vessel interlinked coils.
             The translucent rectangle 
             indicates the cross-section shown in (a) with the $x$ and $y$ 
             directions represented by the green and blue arrows.}
    \label{fig:launcher}
\end{center} \end{figure}

A schematic of the orientation of the launch antenna relative to the plasma is
shown in Fig.~\ref{fig:launcher}. The 2.45 GHz Gaussian 
beam is also depicted, and contours mark  
the first and second harmonics of the EC frequency 
($|\mathbf{B}| = 87.5$ mT and 43.8 mT, respectively).
This cross-section of the non-axisymmetric CNT plasma was selected due to
its ``tokamak-like'' appearance, in the sense that, at that location, 
$|\mathbf{B}|$ decreases monotonically with
the major radius $R$, so that the beam can be launched from the low-field side.
Due to lack of port access in the midplane of the toroidal plasma 
(which is actually vertically oriented in the laboratory frame), 
the antenna enters from a port slightly above the midplane,
and is angled such that its axis intersects the vacuum magnetic axis.

To help select the desired polarization (parallel or perpendicular to
the magnetic field), the magnetic axis is visualized by means of
a hot-cathode \cite{brenner2008} before the microwave discharge. 
Note that the field's pitch angle, when viewed through the waveguide, 
changes by less than $20^\circ$ between the axis and the edge.

\subsection{Diagnostics}
\label{subsect:10kWdiagnostics}

The primary plasma diagnostic used in this work was an array of Langmuir probes 
similar to what was used to diagnose non-neutral
plasmas in CNT.\cite{kremer2006} The probe tips are made of halogen light 
bulbs with the glass removed to expose the tungsten filaments. The tips are
mounted on a ceramic rod which can be moved longitudinally into and out of 
the plasma with an edge-welded bellows drive actuated
by a recently installed stepping motor,  
at a speed of $\approx$ 1 cm/s. The probe enters the plasma in a wide 
cross-section and therefore must move about 30 cm to scan from the edge to
the axis (the average minor radius is 13 cm). For comparison, plasma discharges
may last up to 45 s, limited by the heating of the coils, although
more typical discharges of only 8-10 s are examined here. 
The probe array intersects the plasma between $\phi = 180^\circ$
and $\phi = 215^\circ$, far from the launch antenna which
aims at $\phi=90^\circ$.

Electron temperature and density were determined from probe current-voltage
characteristics $I(V)$, obtained by sweeping the probe bias with a repetition
rate of 200 Hz (that is, every 5 ms).  The steady-state 
measurements reported in this paper were derived from averaging 0.25 - 0.5 s 
of data.

The effective minor radius (flux coordinate) of the probe at each longitudinal
position was determined using an electron beam/phosphor rod technique commonly
used to visualize flux surfaces.\cite{hammond2016,jaenicke1993,lazerson2016}
In this case, the probe itself emitted the electron beam, and the flux surface
images were aligned 
with previous measurements, thereby associating each probe position with a
three-dimensional flux surface. The geometry of these surfaces is well 
understood following a diagnosis of CNT's error fields.\cite{hammond2016}

Electron temperature $T_e$ and density $n_e$ are assumed uniform on the 
flux surfaces. Outside the last closed flux surface (LCFS), however, few 
measurements are available and the extrapolation of $n_e$ to locations close to 
the launch window is subject to uncertainties.  
Models of magnetized plasma sheaths near solid surfaces
predict that a magnetic pre-sheath will extend into the plasma by
a distance given by the sound speed divided by the ion cyclotron
frequency, $c_s/\omega_\mathrm{ci}$.\cite{stangeby2000}
Near the window, this distance is of order 
1 cm for singly ionized Ar, which is the working gas used in these experiments. 
The layer in which the density drops  
has little effect on wave-propagation, due to its thinness compared to the
wavelength and other scales of interest. Additionally, that layer is highly
underdense to O-mode.

\section{Profile measurements: evidence of overdense heating}
\label{sect:profiles}

Microwave plasmas were generated with the heating system described
in Sec.~\ref{subsect:heatingSystem} and diagnosed with
the probes of Sec.~\ref{subsect:10kWdiagnostics},
finding evidence of overdense plasma heating.

Electron temperature and density profiles were obtained for various heating
powers and magnetic field strengths. Each profile was obtained from
a number of discharges realized with the same heating and backfill pressures
parameters, while the probe was scanned through the plasma and 
scrape-off layer. In a typical 8-10 s long discharge, the 
power gradually increased as the magnetron 
warmed up, reaching its flat top at $t$ = 4 s. However, some plasma
parameters were observed to evolve from $t$ = 5 s onwards, especially in
high-power discharges. Therefore, profile mesurements were restricted
to the interval $t$ = 4 - 5 s in each discharge. 
To generate profiles, the probe was radially scanned by $\approx$1 cm 
during that interval, as well as from one discharge to the other.
Tests confirmed the discharge-reproducibility to be well
within the $n_e$ and $T_e$ error bars.

\subsection{Dependence on heating power}
\label{subsect:pscans}

\subsubsection{O-mode launch}\label{subsubsect:OML}

\begin{figure*} \begin{center}
    \includegraphics[width=\textwidth]{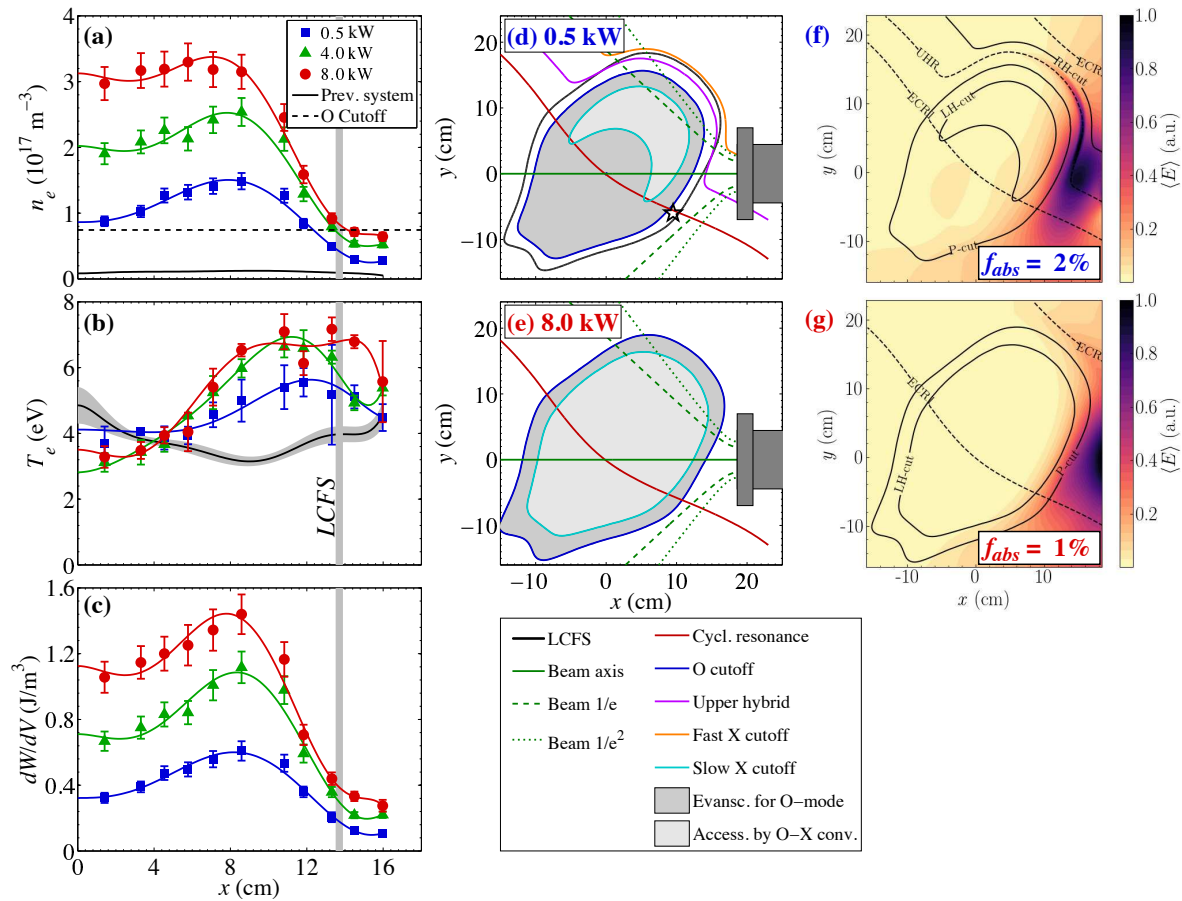}
    \caption{Results for O-mode launch at 
             various power levels.
             (a)-(b) Langmuir probe measurements of $n_e$ and $T_e$,
                     projected on the $x$ axis (as defined in 
                     Fig.~\ref{fig:launcher}), fitted with 
                     7th-order polynomials. The vertical gray line 
                     denotes the LCFS. The horizontal dashed line is the cutoff 
                     density for O-mode. The solid black curve represents
                     a typical $n_e$ profile from 
                     CNT's earlier 1 kW heating system,
                     with $\mathrm{N}_2$ instead of Ar as the working gas.
                 (c) $dW/dV$ data and spline fits.
             (d)-(e) Contours of cutoffs and resonances in the $xy$ plane
                     (Fig.~\ref{fig:launcher})
                     for 0.5 kW and 8.0 kW heating power.
                     Regions overdense to the O-mode are
                     shaded; portions that could be accessible by 
                     O-X conversion are shaded lighter. 
                     The solid, dashed and dotted green lines indicate
                     the axis and two contours of relative intensity (at 
                     $1/e$ and $1/e^2$ of the on-axis values) of
                     the microwave-beam, as it would propagate in vacuum.
                     The star in (d) is a reference for the discussion in 
                     Sec.~\ref{subsect:cyclotronDamping}. The vacuum wavelength
                     is shown in (e).
                     (f)-(g) Contours of time-averaged electric field and
                     percentage of first-pass power absorption determined 
                     by the full-wave code for 0.5 kW and 8.0 kW.}
    \label{fig:o1_pwr_scan}
\end{center}
\end{figure*}

Figs.~\ref{fig:o1_pwr_scan}a-b show $n_e$ and $T_e$ profiles for Argon (Ar)
plasmas 
heated with O-mode-launched waves at different heating powers $P$. 
The profiles were fitted with 7th-order polynomials with the constraint
of $dn_e/dx=dT_e/dx=0$ at $x$=0, for reasons of symmetry.

The 
backfill pressure was $(1.4 \pm 0.2) \times 10^{-5}$ torr in each case. 
Consequently, the mean-free-path of neutrals before being 
ionized was $\gtrapprox 10$ cm. As the plasma minor radius was
$a \simeq$ 13 cm, the neutral density was assumed to uniformly evaluate 
$n_n =  4.5 \times 10^{17}~\mathrm{m}^{-3}$
throughout the plasma, corresponding to the backfill pressure.

For the most part, in the cold plasmas considered the Ar atoms were either
singly ionized or not ionized at all, and there were very few
ions of higher charge. Hence, by quasi-neutrality 
the ion density was $n_i \simeq n_e$. In turn, $n_e$ plotted in 
Fig.~\ref{fig:o1_pwr_scan}b was comparable with the neutral density just
estimated. Consequently, $n_i \approx n_n$, i.e.~the gas was only partly
ionized.
As a consequence, increasing the heating power
did not result solely in heating (higher $T_e$), but also in more ionization
(higher $n_e$), as seen in Figs.~\ref{fig:o1_pwr_scan}a-b. 
Therefore, as a metric of effectiveness of
power-coupling to the plasma, it is convenient to also plot,
in Figs.~\ref{fig:o1_pwr_scan}c, 
\begin{equation}
  \frac{dW}{dV}= \frac{3}{2} n_i k_B T_i + \frac{3}{2} n_e k_B T_e +
                 n_i {\cal E}_\mathrm{ioniz}
  \simeq n_e ( 1.95 k_B T_e +  {\cal E}_\mathrm{ioniz}).  
\end{equation}
Here $k_B$ is the Boltzmann constant and ${\cal E}_\mathrm{ioniz}$ is the
first ionization potential. For Ar it is ${\cal E}_\mathrm{ioniz}$= 15.8 eV,
clearly not negligible in the enery balance of the partly ionized plasmas
of $k_B T_e$= 2-7 eV presented here. 
The quantity $dW/dV$
is the volumetric density of energy $W$ deposited in the plasma, 
expressed in terms of thermal energy of the ions, of the electrons,
and energy expended in the ionization process, per unit volume. The
assumption $T_i\simeq 0.3 T_e$ was made for the ion temperature in
the electron-heated CNT plasma, and discussed in a previous
paper \cite{hammond_pop2017}. 

The profiles of $T_e$ and $dW/dV$ (Fig.~\ref{fig:o1_pwr_scan}b-c)
peaked at the edge, and peaks were found to depend on the heating power $P$,
suggesting power-deposition at the edge. 

Density profiles (Fig.~\ref{fig:o1_pwr_scan}a) show 
that nearly the entire plasma was overdense to O-mode 
propagation, at all power levels. The region of $T_e$ peaking and likely
power deposition was also overdense.

Also shown in Fig.~\ref{fig:o1_pwr_scan}a is the 
$n_e$ profile from an earlier nitrogen plasma produced with a 
rudimentary heating system, consisting of a
1 kW magnetron placed in front of a viewport.
Compared to that, $n_e$ in the new experiments increased by over
an order-of-magnitude, even when using as little as 0.5 kW (in blue in 
Fig.~\ref{fig:o1_pwr_scan}a). This indicates that the new launcher, closer
to the plasma and better aimed, makes a much better use of the injected power,
and deposits a much higher fraction of it into the plasma.
A further improvement could be brought by a focusing, steerable mirror.
Focusing is beneficial in general, and steering of a collimated beam is
beneficial for the O-X-B scheme. 
As for the temperature, $T_e$ also increased with the new launcher, but less
significantly 
(Fig.~\ref{fig:o1_pwr_scan}b). This is due to the plasma not being fully
ionized, as just discussed. Therefore, it is not surprising that
increasing the power coupled to the plasma primarily results
in more ionization, i.e.~higher $n_e$. 

Figs.~\ref{fig:o1_pwr_scan}d-e show contours of cutoffs and resonances in the
$xy$ cross-section defined in Fig.~\ref{fig:launcher}.
The contours were calculated from the 
$n_e$ profiles for O-mode launch
at the lowest and highest power levels. Also shown are the  
launch window and the propagation 
axis and width of the ``vacuum microwave beam''
(i.e., the microwave beam as it would look in the absence of 
plasma). As indicated by the O-mode cutoff curves, both cross-sections are 
almost entirely overdense -thus, evanescent-
to the O-mode. However, some O-X conversion may 
occur and make the lighter-shaded regions partly accessible to the SX mode,
up to a ``turning point'' (not shown) where it bends toward the UHR. 

Finally, shown in Figs.~\ref{fig:o1_pwr_scan}f-g are full-wave simulations
realized with the IPF-FDMC code \cite{koehn2008}.
These simulations will actually be discussed in
Secs.~\ref{sect:fullWave}-\ref{sect:10kW_discussion}, but it is
convenient to plot them here, next to the experimental profiles
(Figs.~\ref{fig:o1_pwr_scan}a-c) and accessibility plots
(Figs.~\ref{fig:o1_pwr_scan}d-e) which they refer to.

\begin{figure*}
\begin{center}
    \includegraphics[width=\textwidth]{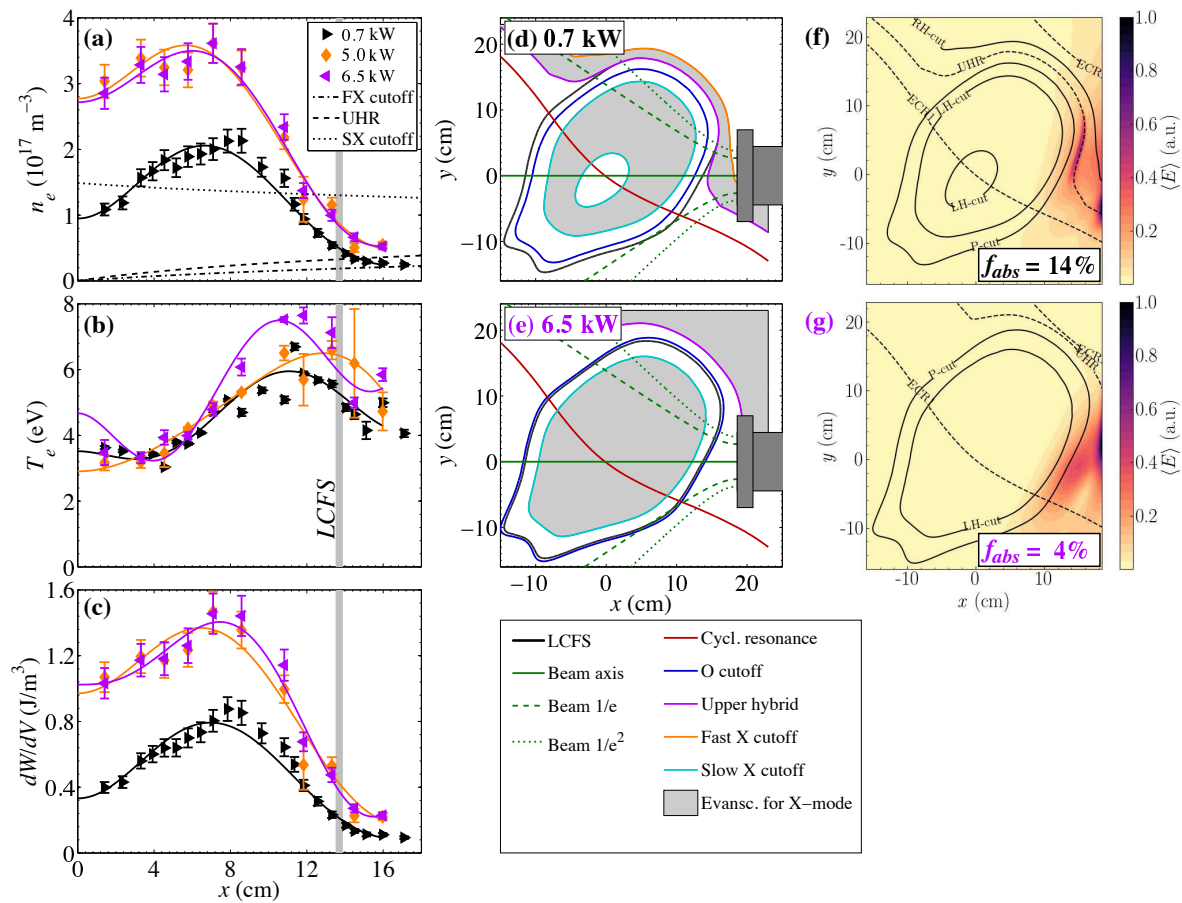}
    \caption{Similar to Fig.~\ref{fig:o1_pwr_scan}, but for 
             X-mode launch. The shaded regions in (d) and (e) indicate
             regions of evanescence for the X-mode.}
    \label{fig:x1_pwr_scan}
\end{center}
\end{figure*}

\subsubsection{X-mode launch}

Fig.~\ref{fig:x1_pwr_scan} presents the results of another power scan,  
for the same Ar gas pressure, but launching an X-mode. 
The shapes and trends in the $n_e$, $T_e$ and
$dW/dV$ profiles are similar to the O-mode case. Again, heating seems 
to occur predominantly at the plasma edge. The low, medium and high 
power levels were obtained using the same stub tunings as in
the O-mode power scan; however, the values of $P$ were 
different. This indicates that X-mode coupling to the plasma differed 
from O-mode coupling. This was expected, due to the different reflectivity of
the plasma to the two modes, which, in turn, is related to the different
locations of the O- and X-mode cutoffs. Cutoffs and evanescent regions
for the two modes are plotted respectively in 
Figs.~\ref{fig:o1_pwr_scan}d-e and \ref{fig:x1_pwr_scan}d-e. 

In generating those figures, 
$n_e$ and $T_e$
were extrapolated outside the LCFS as discussed in 
Sec.~\ref{subsect:10kWdiagnostics}. As a result, contours outside the LCFS
(based on radial extrapolations) are 
less accurate than inside the LCFS, which are based on Langmuir probe
measurements, although at a different toroidal location, far
from the launcher (Fig.~\ref{fig:launcher}b). 
For O-mode launch, this inaccuracy should not impact the analysis 
because $n_e$ is usually well below cutoff outside the LCFS, hence refraction 
of the O-mode is negligible and insensitive to relatively small
$n_e$ variations.
The effect on X-mode propagation might be more significant because 
$n_e$ outside the LCFS is on the order of the 
FX cutoff density. Hence, an error in density 
could switch the medium from overdense to underdense for the X-mode.
Yet, even if the region in front of the launcher becomes evanescent for the
X-mode, as in the case of Fig.~\ref{fig:x1_pwr_scan}d (as opposed to 
Fig.~\ref{fig:x1_pwr_scan}e), finite tunnelling is still possible,
provided the region is not too thick.

\begin{figure} \begin{center}
    \includegraphics[width=0.47\textwidth]{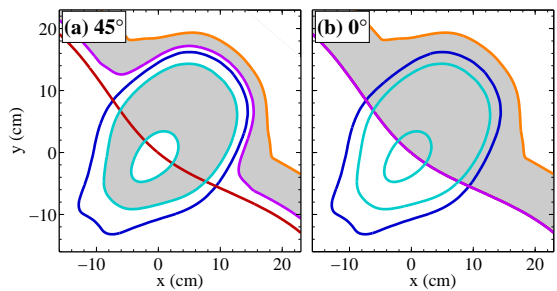}
    \caption{Cutoffs and resonances for the plasma heated with 0.7 kW with
             X-mode launch (Fig.~\ref{fig:x1_pwr_scan}d) calculated assuming
             propagation
             (a) oblique at $45^\circ$ relative to the magnetic field and
             (b) parallel.
             Line and fill colors are as defined in the legend of 
             Fig.~\ref{fig:x1_pwr_scan}. In the parallel case, the shaded
             regions are evanescent to the right-handed (R) mode.}
    \label{fig:angle_compare}
\end{center} \end{figure}

It should also be noted that, for simplicity, the cutoff and resonant 
layers in Figs.~\ref{fig:x1_pwr_scan}d-g were determined assuming 
propagation at $90^\circ$ relative to the magnetic field. This is not 
the case everywhere, due to the incident beam being angularly broad
(due to the low frequency) and 
the curvature of $\mathbf{B}$ being appreciable over the transverse size of the
beam (due to CNT's low aspect ratio and to the beam being broad).
As a consequence, in different locations the propagation vector $\mathbf{k}$ 
forms a different angle with $\mathbf{B}$, and different from $90^\circ$. 
The evanescent region between the UHR and FX cutoff
varies accordingly (Fig.~\ref{fig:angle_compare}) and becomes thicker 
for shallower incidence. 
This, combined with the larger distance traveled across the
layer for more oblique angles, reduces the X-mode tunneling efficiency
for grazing incidence.

\begin{figure} \begin{center}
    \includegraphics[width=0.47\textwidth]{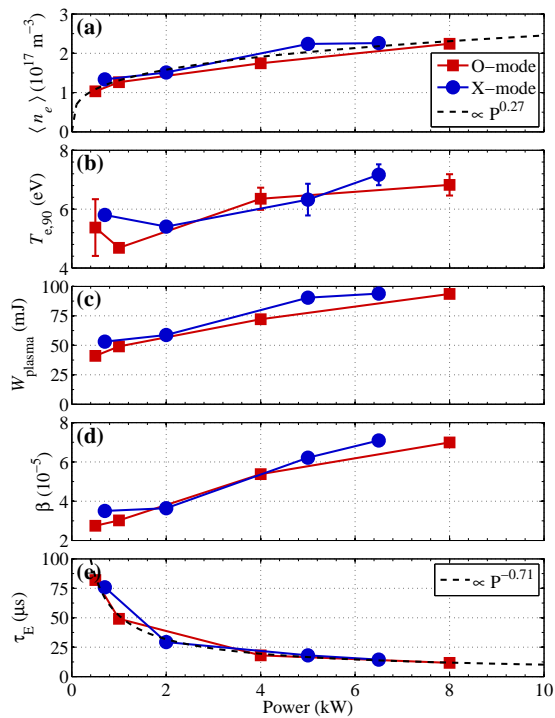}
    \caption{Dependence of plasma parameters on heating power.  
             (a) Volume-averaged electron density, with a fit of a power law, 
             (b) Electron temperature on the flux surface at 90\% of the 
                 effective minor radius, 
             (c) Plasma energy,
             (d) Volume-averaged $\beta$,
             (e) Energy confinement time, with a fit of a power law.}
    \label{fig:pwr_scan_1d}
\end{center} \end{figure}

\subsubsection{Dependence of global parameters on heating power}

The density and temperature profiles in 
Figs.~\ref{fig:o1_pwr_scan}-\ref{fig:x1_pwr_scan} increased with the
injected heating power, but maintained their shapes nearly
unaltered. Therefore, their dependence on power can be summarized
by only plotting the volume-averaged density $\langle n_e \rangle$ and
edge temperature $T_{e,90}$, as in Fig.~\ref{fig:pwr_scan_1d}a-b.
Here $\langle n_e \rangle = (1/V)\int n_e(\rho) dV$, where $n_e(\rho)$ is the
fitted experimental profile,
which is a function of the flux surface coordinate $\rho$. The temperature 
$T_{e,90}$ is the fitted $T_e$ evaluated at 90\% of the effective minor
radius. That was the approximate location where $T_e$ tended to reach
its maximum. Hence, $T_{e,90}$ is effectively the peak temperature, 
or close to the peak temperature.

The data points in 
Fig.~\ref{fig:pwr_scan_1d}a-b were all obtained using $|\mathbf{B}|$ = 88 mT.
They are a combination of all data from 
Figs.~\ref{fig:o1_pwr_scan}-\ref{fig:x1_pwr_scan} as well as the 88 mT data from
a $|\mathbf{B}|$ scan 
to be presented in Figs.~\ref{fig:o1_modb_scan}-\ref{fig:x1_modb_scan} and 
discussed in Sec.~\ref{subsect:bscans}.

The temperature increases with the heating power $P$, as expected
(Fig.~\ref{fig:pwr_scan_1d}b).
The density also increases with $P$, approximately like $P^{0.27}$
(Fig.~\ref{fig:pwr_scan_1d}a). 
This $\langle n_e \rangle$ increase is due to
the plasma not being fully ionized, as discussed in Sec.~\ref{subsubsect:OML}. 

The energy stored in the plasma and the volume-averaged plasma beta 
were calculated as 
\begin{eqnarray}
  W_\mathrm{plasma} & = & \int n_e(\rho)
  [1.95 k_B T_e(\rho) +{\cal E}_\mathrm{ioniz} ] dV,\\
    \beta & = & \frac{1}{V}
    \int 
    \frac{1.95 n_e(\rho) k_B T_e(\rho)}
    {|\mathbf{B}|^2/2\mu_0} dV.
\end{eqnarray}
As expected, both quantities increase with heating power
(Fig.~\ref{fig:pwr_scan_1d}c-d). 

Note that the cold plasmas under consideration ($T_e<$ 10 eV) 
suffer from large losses from line radiation \cite{tanga1986}. 
It is well-known, however, that higher heating power can overcome this 
``radiative barrier'' and, by reduced line-radiation losses, establish 
a more favorable power balance, resulting in significantly
higher temperatures and, therefore, values of $\beta$ \cite{hammond_pop2017}.

Finally, the energy confinement time (Fig.~\ref{fig:pwr_scan_1d}e)
was estimated as $\tau_E=W_\mathrm{plasma}/P$. Note that $P$ is
an upper bound for the coupled power (Sec.~\ref{subsect:heatingSystem}).
Therefore,
the plotted $\tau_E$ is really a lower bound for the confinement time.

Energy confinement is found to decrease with $P$ like $P^{-0.71}$ 
(Fig.~\ref{fig:pwr_scan_1d}e).
At first sight this seems in good agreement with 
the International Stellarator Scaling ISS04 \cite{yamada2005} 
\begin{equation}
  \tau_E =
  0.134 f_{ren} a^{2.28} R^{0.64} P^{-0.61}
  \bar{n}_e^{0.54} B^{0.84} \iotabar_{2/3}^{0.41}
  \label{eqn:tauE}, 
\end{equation}
where $f_{ren}$ is a device-specific parameter, and
$\iotabar_{2/3}$ denotes the rotational transform
evaluated at two thirds the minor radius of the LCFS.
However, the scan in Fig.~\ref{fig:pwr_scan_1d} is not a pure scan of $P$,
``all the rest remaining equal''. Rather, $\langle n_e \rangle$ and thus
the line-averaged density
$\bar{n}_e$ are also varying with $P$. Secondly, and most importantly,
Eq.~\ref{eqn:tauE} fitted data from fully ionized stellarator plasmas,
where the confinement degradation was probably governed by transport
physics in plasmas made non-Maxwellian by microwave, radio-frequency and
neutral-beam heating. 

By contrast, power-balance (hence, energy-confinement) in the
partly ionized plasmas considered here suffers from severe radiative losses,
as mentioned above. 
One driver of radiative losses, electron-ion recombination, increases roughly
in proportion with $n_e$ and $n_i$ and thus in proportion with $n_e^2$ in a 
quasineutral plasma. Hence, it is not surprising that $\tau_E$ decreases
(Fig.~\ref{fig:pwr_scan_1d}e) as  
$\langle n_e \rangle$ increases (Fig.~\ref{fig:pwr_scan_1d}a).
Indirectly, this could also explain why $\tau_E$ decreases with $P$
(due to the correlation between $n_e$ and $P$). 

Future power upgrades are expected to yield temperatures in excess of the
radiation barrier, $T_e >$ 50 eV, at
which radiative losses will diminish. As a result 
$\tau_E$ is expected to increase dramatically 
and not to follow the rough $\tau_E \propto n_e^{-2}$ and
$\tau_E \propto P^{-0.71}$ scalings just
discussed, but rather the international scaling in Eq.~\ref{eqn:tauE} 
with an adequate $f_{ren}$, probably $f_{ren}=0.21 \pm 0.04$,
as argued in Ref.\cite{hammond_pop2017}. 
For the same reasons, much higher values of 
$\tau_E$ are expected, compared to Fig.~\ref{fig:pwr_scan_1d}e, once the
radiative barrier is exceeded \cite{hammond_pop2017}.

\subsection{Dependence on magnetic field}
\label{subsect:bscans}

A series of experiments was also conducted in which the heating power was 
kept constant but the strength of the magnetic field, $|\mathbf{B}|$, 
was varied. The purpose of this was to move the 
cyclotron resonance and, to a lesser extent, the UHR and the X-mode cutoffs,
but not the O-mode cutoff. These changes affected the efficiency of
candidate mode-conversion and heating mechanisms. 
Therefore, the $|\mathbf{B}|$ scan had the promise of helping to isolate 
which mechanisms were taking place in the experiments. 

\begin{figure*}[t]
\begin{center}
    \includegraphics[width=\textwidth]{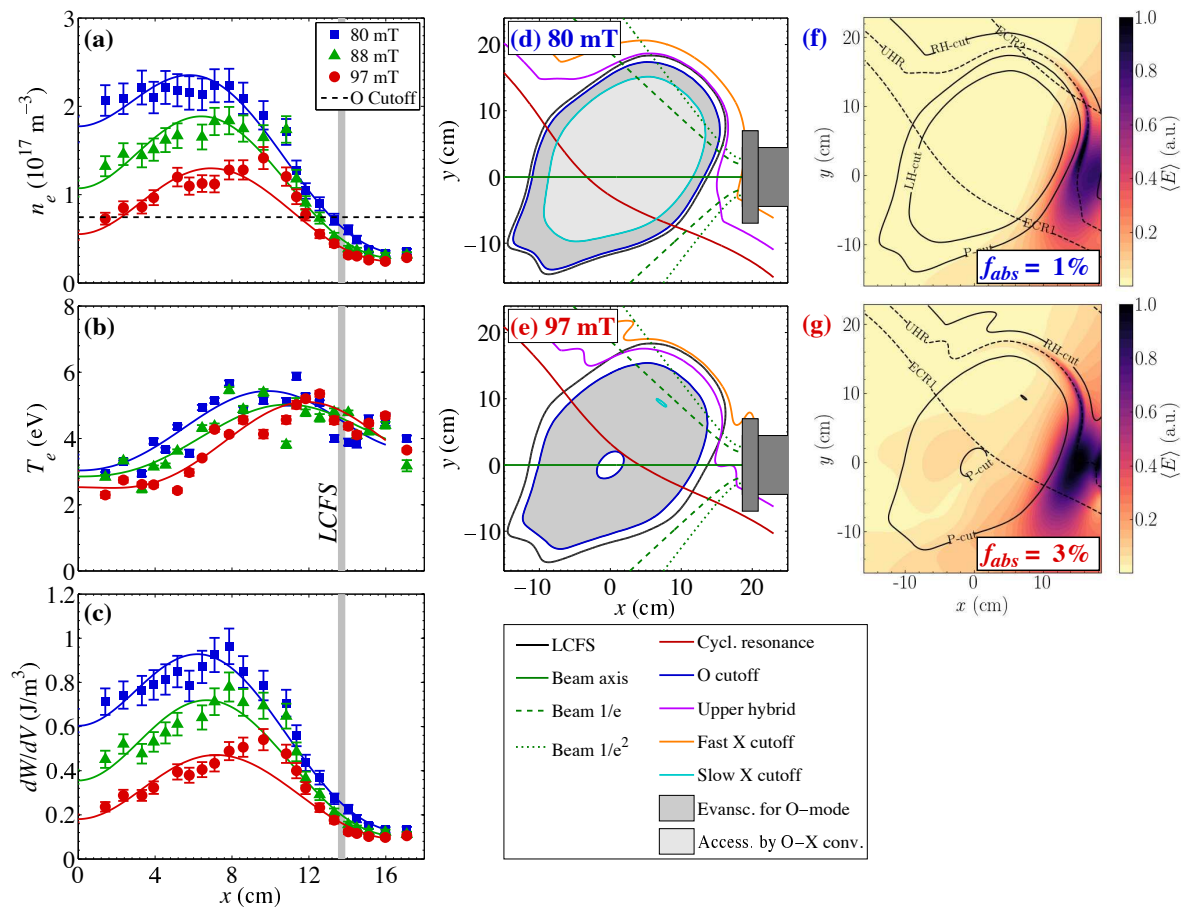}
    \caption{Similar to Fig.~\ref{fig:o1_pwr_scan}, but for the scan of 
             magnetic field strengths in the case of O-mode launch. Each plasma 
             coupled to 1 kW of heating power.}
    \label{fig:o1_modb_scan}
\end{center}
\end{figure*}

\begin{figure*}[t]
\begin{center}
    \includegraphics[width=\textwidth]{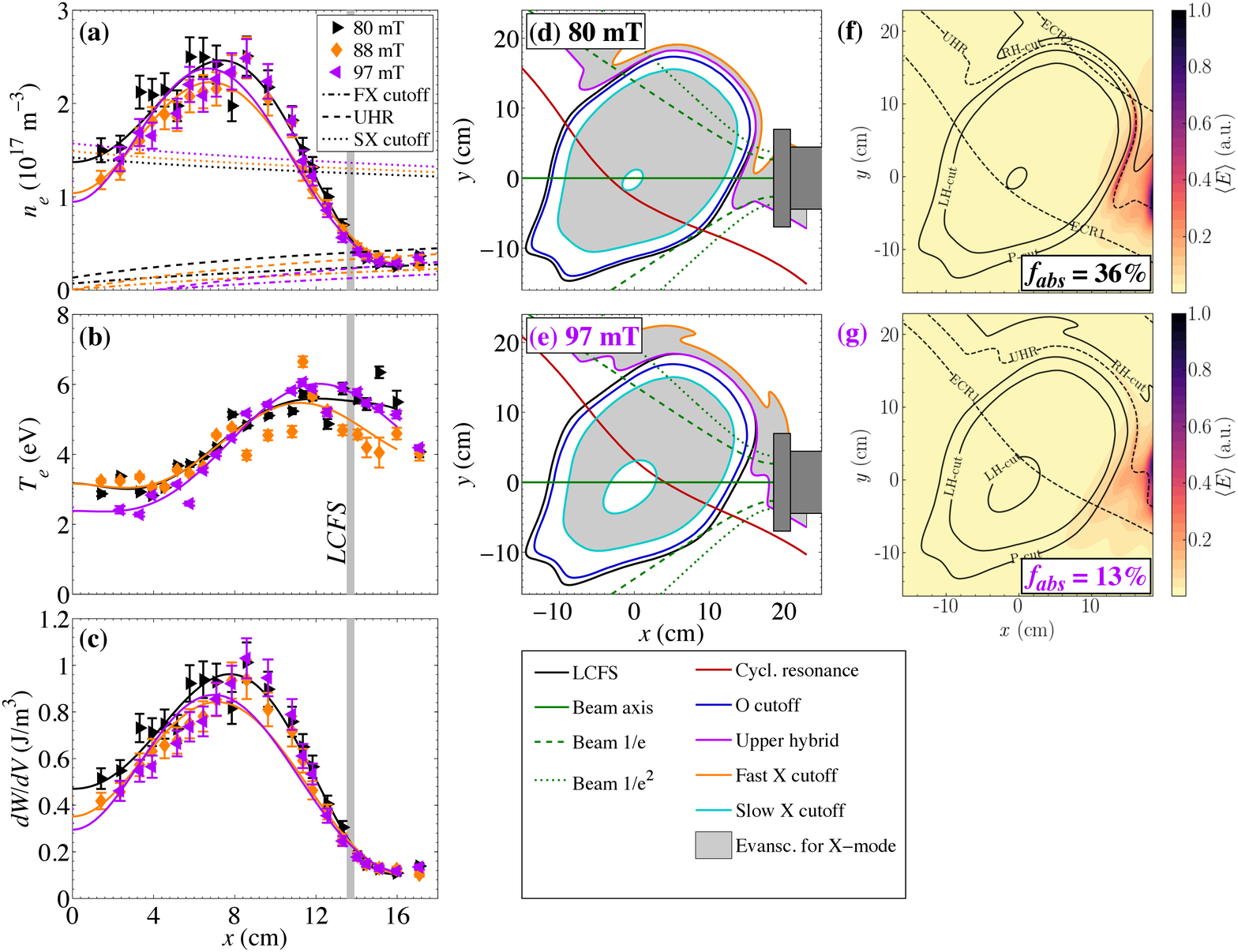}
    \caption{Similar to Fig.~\ref{fig:o1_modb_scan}, but for X-mode launch. 
             Each plasma coupled to 1 kW of heating power, except for the
             $B = 88$ mT case, which coupled to 2 kW.}
    \label{fig:x1_modb_scan}
\end{center}
\end{figure*}

The intermediate $|\mathbf{B}|$ was the same as for the power
scans in Sec.~\ref{subsect:pscans}. The higher and lower $|{\bf B}|$ values 
are the highest and lowest permissible values such that 
neither the fundamental nor the second harmonic resonant surfaces intersect the 
launch window (similar to Fig.~\ref{fig:launcher}a). 
This is to avoid damage to the window. The values of
$|\mathbf{B}|$ given in the legends of
Figs.~\ref{fig:o1_modb_scan}-\ref{fig:x1_modb_scan} 
are evaluated at the origin of the $xy$ plane. 

The stub-tuning was the same for all the experiments in this scan. 
This resulted in net power 
coupling of $1.0 \pm 0.2$ kW in all cases except X-mode launch at 88 mT, 
which had a net coupling of about 2 kW. 

The results are shown in Fig.~\ref{fig:o1_modb_scan} for O-mode launch
and in Fig.~\ref{fig:x1_modb_scan} for X-mode launch. 
For O-mode launch, the $n_e$ and $T_e$ profiles exhibit the greatest
sensitivity to $|\mathbf{B}|$. 
Changes of profile shape were also observed, with both $n_e$ and $T_e$ becoming
increasingly hollow as $|\mathbf{B}|$ increased, and the core density
being halved. 

By contrast, for X-mode launch the profiles exhibited little to no variation
with $|\mathbf{B}|$.


\section{Full-wave modeling}
\label{sect:fullWave}

In order to predict and interpret the experimental results, 
the interactions between injected microwaves and the CNT plasma
was modeled by means of the full-wave, finite-difference time-domain  
code IPF-FDMC.\cite{koehn2008} 
The code solves Maxwell's equations coupled with the fluid 
equation of motion for electrons 
in a nonuniform magnetized plasma.
As a result, it accounts for effects such as the 
O-X mode conversion and tunneling of the X-mode through the evanescent region
between the UHR and fast X cutoff. 

A damping term invoking a collision frequency $\nu$ in the range 
$10^{-5} < \nu/\omega < 10^{-3}$ is assumed, which damps the 
slow X-mode in the vicinity of the UHR\cite{koehn2010}.
Here $\omega$ is the wave frequency. The total damped 
power is not sensitive to the value of $\nu$ within this range.
As it will be discussed in
Sec.~\ref{subsect:collAbs}, collisional damping is expected to 
dominate over conversion to EBWs, in the CNT experiments presented here.
No cyclotron damping is accounted for in these simulations, but is
expected to be negligible anyway (Sec.~\ref{subsect:cyclotronDamping}).

Full-wave calculations were performed  
for all values of $P$, $|{\bf B}|$ and polarizations
in the experimental scans of Sec.\ref{sect:profiles}.
Selected cases (maximum and minimum $P$ and $|{\bf B}|$),
corresponding to panels d and e
of Figs.~\ref{fig:o1_pwr_scan}-\ref{fig:x1_pwr_scan} and 
\ref{fig:o1_modb_scan}-\ref{fig:x1_modb_scan}, are plotted respectively
in panels f and g of the same figures. 
Shown are the contours of steady-state, time-averaged amplitude of the
microwave electric field, $\langle|\mathbf{E}|\rangle$, normalized to its
peak value (which changes from one contour plot to the other). 
The $x,y$ profiles of $n_e$ and $T_e$ adopted in the simulations
are based on the fits of the experimental profiles in panels a and b.

The launcher is located on the right of the computational domain and injects
linearly polarized waves with a Gaussian distribution of intensity.
This is a realistic model of the experiment
(Fig.~\ref{fig:waveguide_schematic}c and \ref{fig:launcher}a). 
Perfectly absorbing conditions are adopted at all other boundaries, unless
noted otherwise.  
Hence, the plotted $\langle|\mathbf{E}|\rangle$ refers to
the first pass of microwaves but neglects 
additional power re-impinging on the plasma after reflections off the
vessel walls or internal coils.  
The percentages of first-pass power absorption, primarily
due to collisional damping, are indicated in panels f and g.

In each case in which the UHR is visible, enhancement of 
$\langle|\mathbf{E}|\rangle$ is noticeable in a narrow region (much narrower
than the vacuum wavelength) close to the UHR. Only the X-mode is sensitive
to this resonance, but enhancement is observed for nominal O-mode launch as
well. This is ascribed to a combination of O-X mode conversion and
lack of modal purity: as discussed above, neither in the experiment nor in
the modeling are the linear polarizations injected 
pure O- or X-polarizations, which would be elliptical. This implies
that some X-mode is injected during nominal O-mode launch. 

Field enhancement near the UHR favors collisional damping. 
In some cases this can account for as much as 36\% absorption in a single
pass (Fig.~\ref{fig:x1_modb_scan}f).
In other cases, only 1\% of power is absorbed at the first transit
(Figs.~\ref{fig:o1_pwr_scan}g and \ref{fig:o1_modb_scan}f). This is due to the
recessed location of the UHR, such that, in those cases,
the beam does not encounter the UHR in its first transit.

\subsection{Modeling multiple reflections}
\label{subsect:modelMultiRefl}

\begin{figure} \begin{center}
    \includegraphics[width=0.47\textwidth]{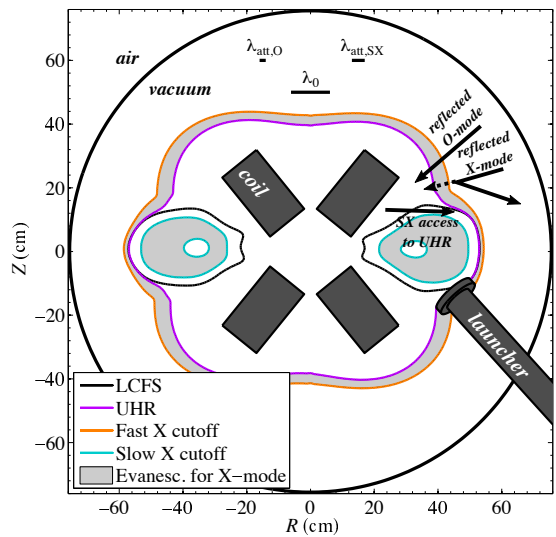}
    \caption{Schematic of how the X-mode may access the UHR from the 
             high-field side. O-waves (originating from the launcher or 
             wall reflections) may pass undisturbed through regions evanescent 
             to X-mode (gray), reflect as X-waves off the in-vessel coils, and
             reach the UHR. In addition, X-waves from the launcher or 
             wall reflections may partially transmit through the outermost
             evanescent layer and reach the UHR either directly or after
             reflecting off the coils.
             Denser plasmas make the evanescent regions thicker, thus 
             restricting wave access. Shown near the top of the diagram are
             references for $\lambda_0$ and the attenuation lengths
             $\lambda_\mathrm{att,O}$ and $\lambda_\mathrm{att,SX}$ for the O-
             and slow X-mode for $n_e = 1.5\times10^{17}~\mathrm{m}^{-3}$ and
             $|\mathbf{B}| = 70$ mT. Note that the launcher is actually located 
             above the cross-sectional plane shown here and does not intersect 
             the LCFS.}
    \label{fig:sx_heating_access}
\end{center} \end{figure}

In the experiment,
the unabsorbed power is reflected by the metallic walls of the vacuum vessel
and by the metal cases of the in-vessel coils, and effectively re-injected
in the plasma. 
To study this effect, the calculations of
Figs.~\ref{fig:o1_modb_scan}g and \ref{fig:x1_modb_scan}g  
were repeated with perfect electrical conductors lined on the left
and bottom edges of the computational domain. The conductors simulate the 
in-vessel coils placed at the approximate same locations,
although with an inclination of 78$^\circ$ (or, equivalently, 102$^\circ$)
relative to each other (Fig.~\ref{fig:sx_heating_access}).

\begin{figure}[t]
\begin{center}
    \includegraphics[width=0.47\textwidth]{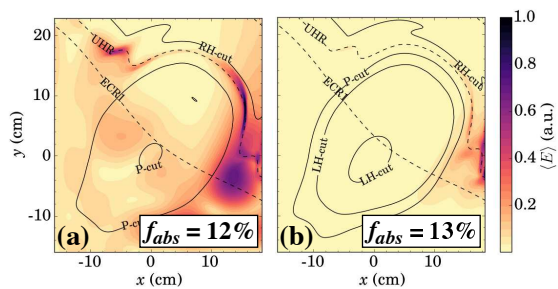}
    \caption{Full-wave simulations for
      (a) O- and (b) X-mode launch from the launcher located on the right of
      the computational domain, similar to
      Fig.~\ref{fig:o1_modb_scan}g and \ref{fig:x1_modb_scan}g, but in the
      presence of reflective boundaries on the left and bottom, to
      simulate waves incident on the plasma 
      after reflections off the in-vessel coils. Note the field
      enhancement in the upper left of
      panel a, compared to Fig.~\ref{fig:o1_modb_scan}g.}
    \label{fig:reflEdge}
\end{center} \end{figure}

For O-mode launch, this results in field-enhancement and absorption
in a new location (upper left of Fig.~\ref{fig:reflEdge}a),
compared with Fig.~\ref{fig:o1_modb_scan}g.
The field pattern for X-mode launch (Fig.~\ref{fig:reflEdge}b),
on the other hand, does not change significantly with respect to
the same case with fully absorbing boundaries
(Fig.~\ref{fig:x1_modb_scan}g). In brief, reflections may or may not
contribute to field-enhancement and absorption, depending on geometry
and polarization. The same is expected of wall reflections.

\begin{figure*}[t]
\begin{center}
    \includegraphics[width=\textwidth]{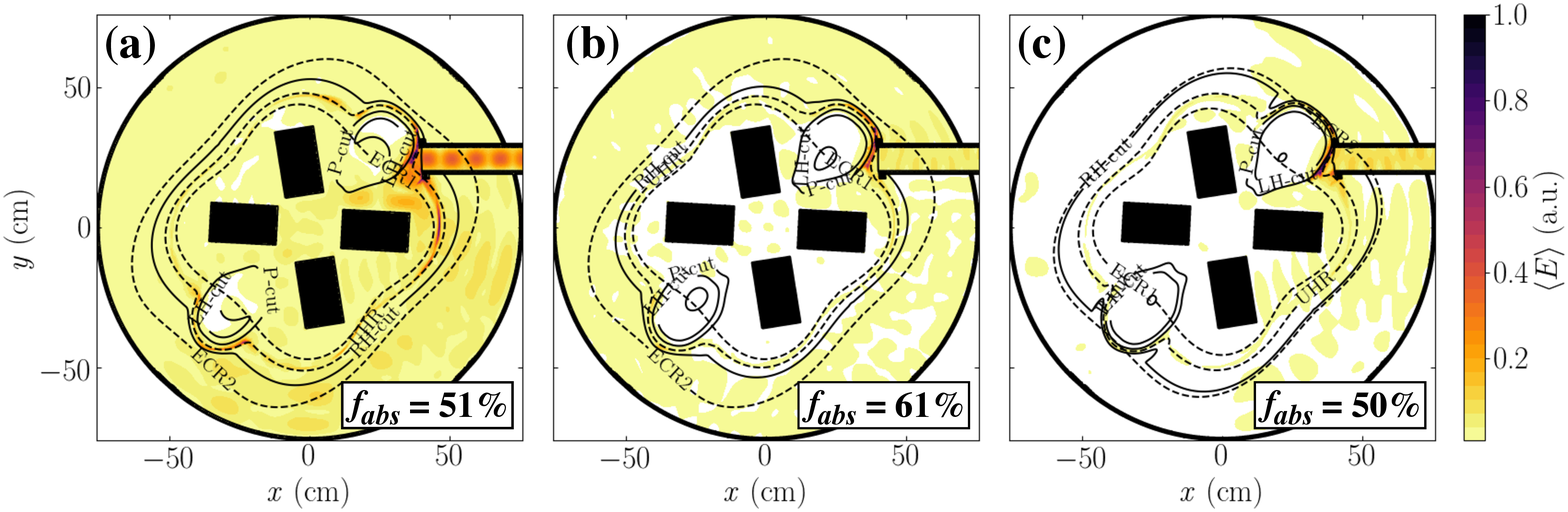}
    \caption{(a) Full-wave two-dimensional 
      simulation of the steady-state, time-averaged wave electric field for the 
      low-power (0.5 kW), medium field (87.5 mT), O-mode launch case of
      Fig.~\ref{fig:o1_pwr_scan}f, but in a larger computational
      domain, similar to Fig.~\ref{fig:sx_heating_access}. Note the
      wave enhancement at the UHR at approximately $x=-$30 cm, $y=-$30 cm,
      a location that waves reach after multiple wall reflections.
      Also note the standing wave in the waveguide, due to partial
      reflection at the O-mode cutoff. (b) Like (a), but for X-mode launch.
      Compare with first pass in Fig.~\ref{fig:x1_pwr_scan}f. 
      (c) Like (b), but for low power (1 kW) X-mode launch in low field plasma
      (80 mT). Compare with first pass in Fig.~\ref{fig:x1_modb_scan}f. 
      Note that, in each contour plot,  
      $\langle|\mathbf{E}|\rangle$ is normalized to its maximum value in
      that particular plot. 
      }
    \label{fig:fullDevice}
\end{center}
\end{figure*}

These effects were integrated in two-dimensional (2D)
full-device simulations. The computational domain is now a  
cross-section of the entire vessel, which the plasma intersects twice.
One result in this larger domain, for the case of O-mode at low power
with B = 87.5 mT, is shown in Fig.~\ref{fig:fullDevice}a.
Significant field enhancement is visible at the lower left of the figure.
There is no direct line of sight from the launcher to that toroidally
remote location, which the wave reaches after multiple
reflections off the walls and coils.

As a result, the fraction of absorbed power increases from 2\%
for first-pass (Fig.~\ref{fig:o1_pwr_scan}f) to 51\%
when multiple reflections are taken into account 
(Fig.~\ref{fig:fullDevice}a). The remaining 49\% of the
power was reflected back into the waveguide, as 
evidenced by the standing-wave pattern in the launch antenna in the
upper-right corner of Fig.~\ref{fig:fullDevice}a. 
Other standing waves can also be noticed, between the plasma and the wall. 
All twelve experimental combinations of $P$, $|{\bf B}|$ and polarization
of Figs.~\ref{fig:o1_pwr_scan}-\ref{fig:x1_pwr_scan} and 
\ref{fig:o1_modb_scan}-\ref{fig:x1_modb_scan} were numerically modeled. 
For brevity, contours of $\langle|\mathbf{E}|\rangle$ are only shown 
for eight cases in the restricted domain with absorbing boundaries
(panels f-g of the said figures) and for three cases in the enlarged domain,
with reflections included (Fig.~\ref{fig:fullDevice}). 
However, significant increases of fractional absorption were observed in all
twelve cases. Barring three exceptions at 15\%, 29\% and 39\%,
the fraction of power
absorbed amounted to 50-80\%. These calculations underscore the 
crucial role of reflections in coupling the microwave
power to the CNT plasma in these experiments.

Even higher absorption percentages are expected in 3D simulations. This is
because the fractional surface of the launcher and other unshielded ports 
(relative to the total surface of the wall) is even smaller than the 
fractional arc subtended by the launcher 
(relative to the total circumference of the wall) in the 2D problem.
Therefore, on average in 3D the beam will experience more
reflections before impinging on a port and abandoning the vessel. 
Consequently, it will cross the plasma a higher number of times,
and deposit more power. 

Resistive losses at the wall are neglected because the
reflectivity of stainless steel to
2.45 GHz microwaves exceeds 99.9 \% \cite{dressel2011}.

It is well known that an incident O-mode (or X-mode) can be partly
reflected by smooth metallic surfaces as X-mode (or O-mode).
This property is known as depolarization or polarization mixing, and
in the fusion literature it is often referred to as polarization scrambling 
\cite{albajar2005}. When using reflective boundaries, as in
Figs.~\ref{fig:reflEdge} and \ref{fig:fullDevice},  
IPF-FDMC naturally accounts for this effect.

Corrugated surfaces can cause additional scrambling
\cite{BeckSpizz},
to the point that depolarization is often used as a measure of surface
roughness \cite{liu2015}.
This is because waves of different polarizations tend to be reflected at the
corrugations' crests and troughs, respectively. Said otherwise,
realistic features, unless much smaller than $\sim\lambda/10$,
introduce phase-shifts that affect polarization.
In the present case, thanks to the
large wavelength $\lambda$, it was sufficient to model the CNT vessel
and coils with $\sim$1 cm accuracy to take this effect into account as well.

%

Yet, the amount of polarization scrambling observed in the simulations was
relatively modest. For instance, the X wave launched in
Fig.~\ref{fig:fullDevice}b remains fairly purely polarized in X-mode
even after reflections, as indicated by the fact that it barely penetrates
beyond the right-handed FX cutoff. 

Finally note that if single-pass absorption is high
(Fig.~\ref{fig:x1_modb_scan}f),  
very few reflections occur before most of the power is completely
absorbed by the plasma or reflected back in
the waveguide, as illustrated in Fig.~\ref{fig:fullDevice}c.

\section{Discussion: mode conversion, UHR accessibity, and heating mechanisms}
\label{sect:10kW_discussion}

In the following Subsections various mode-conversion and wave-damping
mechanisms are examined one by one, and invoked or discarded as
possible explanations of the experimental results presented in
Figs.~\ref{fig:o1_pwr_scan}-\ref{fig:x1_modb_scan}, on the basis of
analythical or numerical arguments.

\subsection{SX-B mode conversion}
\label{subsect:SX-B}

The SX-B mode conversion is not possible in the first pass, 
because the beam is launched from the low-field side. 
However, all plasmas discussed here were overdense both to the O- and X-mode 
(see for example $n_e$ profiles, cutoff layers and shaded regions
in panels a, d and e in Figs.~\ref{fig:o1_pwr_scan}-\ref{fig:x1_pwr_scan} and 
\ref{fig:o1_modb_scan}-\ref{fig:x1_modb_scan}). 
In fact, the beams encounter the O and FX cutoff layer within 15 cm of the 
launch window. 
Consequently, part of the incident power is reflected by the plasma
and experiences multiple reflections off the chamber walls, off the
in-vessel coils (coated in metal jackets), and off the plasma. 
At the wall and at the coils, ``polarization
scrambling'' occurs \cite{albajar2005}, causing incident O-mode to be 
reflected as X-mode and vice versa. Some of the reflected X-mode (whether
it conserved its original polarization or it originated from an O-mode launch) 
may access the UHR from the 
high-field side, resulting in some SX-B conversion 
(Fig.~\ref{fig:sx_heating_access}).

It should be noted that 
much of the core plasma is overdense to slow X waves 
(see, for example, Fig.~\ref{fig:x1_pwr_scan}d-e). This creates a
``bottleneck'' narrower than $\lambda_0$ between the overdense core and the UHR.
Such bottleneck restricts wave access to the UHR inside or near the LCFS
(again, see Fig.~\ref{fig:x1_pwr_scan}d-e, or see 
arrow in Fig.~\ref{fig:sx_heating_access}), thereby 
partly inhibiting the SX-B conversion. 
Accessibility would improve if $n_e$ was lower, due to the smaller
overdense core. 
However, the 
fact that the case of least-restricted access was also the case with the
lowest core density and temperature (Fig.~\ref{fig:o1_modb_scan}) indicates
that SX-B conversion does not play a significant role in heating the plasma.

\subsection{FX-B conversion}
\label{subsect:FX-B}

First-pass FX-B conversion should be possible. Its efficiency depends on the 
thickness $\Delta x$ of the layer between the FX cutoff and the 
UHR.\cite{budden1961} The power transmission $T^2$ through this layer is
approximately $\exp(-\pi\eta)$, where 
$\eta = |2\pi \Delta x/\lambda_0|$ is the Budden factor
(Sec.~13.5 of Ref.\cite{stix1992}). 

For example,
the evanescent layer in front of the launcher in Fig.~\ref{fig:x1_modb_scan}e 
is as thin as $\Delta x$ = 0.5 cm, along the beam axis. Compared with the vacuum
wavelength $\lambda_0$ = 12.2 cm, this gives $T^2$ = 45\%, along the beam axis.
That is, nearly 
half the power injected in X-mode along that ray
can tunnel through the evanescent region and reach the UHR.

It should also be noted, however, that, while thin in certain locations
(for example, along the beam axis in Fig.~\ref{fig:x1_modb_scan}e), 
the evanescent layer can be thicker elsewhere 
(for example, at higher $y$ in Fig.~\ref{fig:x1_modb_scan}e).
As a result, if we model the beam as a bundle of rays,
some rays have to tunnel through thicker regions, with reduced transmissivity.
Density can also fluctuate as a function of time, or from one discharge to
another, and if $n_e$ decreases by just 25\% outside the LCFS,
$\Delta x$ increases from 0.5 cm to 5 cm in the case considered,
giving $T^2 < 0.1\%$.

In conclusion, if the FX cutoff and the UHR layers lie in front of the
launcher, and the region in between is sufficiently thin, as is the case
in Fig.~\ref{fig:x1_modb_scan}e, significant power reaches the UHR directly 
at the first transit, without assistance from wall reflections. 
At the UHR, it can either convert to EBWs or, most likely,
be collisionally damped, as discussed below.
If, instead, the region is thick (for instance, due to low $n_e$),
as is the case in Fig.~\ref{fig:x1_pwr_scan}d or \ref{fig:x1_modb_scan}d,
only a small fraction of X-mode power penetrates it, at every single transit. 
Finally, for sufficiently high $n_e$ the FX cutoff and UHR lie ``behind'' the
launcher, there is no evanescent layer to penetrate, and the beam
impinges directly on the overdense core, as in in Fig.~\ref{fig:x1_pwr_scan}e.

\subsection{O-X-B conversion}
\label{subsect:O-X-B}

The CNT launch antenna emits an angularly broad beam. If treated as a bundle
of rays, not all rays can simultaneously have the special direction
for the O-X conversion to occurr. 
This aspect can be improved in the future by introducing a refocusing, steerable
mirror to collimate the O-mode beam and aim it at the O-mode cutoff with the
optimal angle. 

In addition, the CNT launch antenna can only output linearly polarized waves,
whereas a pure O-mode propagating obliquely to $\mathbf B$ would be
elliptically polarized, which will require the introduction of a $\lambda/4$
phase-shifter. 

Nevertheless, finite O-X coupling is possible in CNT due to its low
$L_n/\lambda_0$ at the location of the O-mode cutoff (0.3 to 
0.4 for the O-mode profiles shown here). Here, $L_n = n/(dn/dx)$ is the
density lengthscale. For such steep density gradients -relatively speaking-,
the tolerance on the optimal angle for O-X conversion is
known to relax.\cite{mjohlus1984,igami2006}

\subsection{Electron cyclotron damping of O- and X-mode}
\label{subsect:cyclotronDamping}

As seen in panels d-e of Figs.~\ref{fig:o1_pwr_scan}-\ref{fig:x1_pwr_scan} and
\ref{fig:o1_modb_scan}-\ref{fig:x1_modb_scan}, 
portions of the EC resonance are directly accessible to
the microwave beam, without mode conversions,
because the plasma is locally underdense.  
This is for example the case of the point marked by the star symbol in 
Fig.~\ref{fig:o1_pwr_scan}d at coordinates (9.5 cm, -6 cm). In principle, 
cyclotron damping of electromagnetic 
O- and X-waves would be possible in these areas, provided that the optical
depth $\tau$ is sufficient.
For finite density and oblique propagation, $\tau$ is given in Table XII 
of Ref.~\cite{bornatici1983}: 

\begin{eqnarray}
\tau_\mathrm{O1} &= \pi^2 N 
    \left(\frac{\omega_\mathrm{pe}}{\omega_\mathrm{ce}}\right)^2
    \left(\frac{v_t}{c}\right)^2
    \frac{\left(1 + 2\cos^2\theta\right)^2\sin^4\theta}
         {\left(1 + \cos^2\theta\right)^3}
    \frac{L_B}{\lambda_0} \\
\tau_\mathrm{X1} &= \pi^2 N^5
    \left(1 + \frac{\omega_\mathrm{pe}^2}{\omega_\mathrm{ce}^2}\right)^2
    \left(\frac{\omega_\mathrm{ce}}{\omega_\mathrm{pe}}\right)^2
    \left(\frac{v_t}{c}\right)^2
    \cos^2\theta \frac{L_B}{\lambda_0}
\end{eqnarray}

\noindent Here $N$ is the index of refraction, $\theta$ is the angle between
the propagation vector and $\mathbf{B}$, $v_t$ is the electron thermal 
velocity, and $L_B = |\mathbf{B}|/(d|\mathbf{B}|/dx)$ is the scale length of 
variation in $|\mathbf{B}|$ along the beam trajectory. The O1 subscript
denotes first-harmonic O-mode; X1 denotes first-harmonic X-mode. 

Fusion plasmas are typically optically thick to first-harmonic O-mode, thanks
to their high $T_e$. In CNT, though,  
${v_t}^2$ is three orders of magnitude lower. Additionally, due to the
low $|\mathbf{B}|$, hence low $\omega_text{ce}$, thus low wave frequency, 
$\lambda_0$ is nearly two orders of magnitude longer.
As a consequence, $\tau_\mathrm{O1} < 10^{-4}$ and $\tau_\mathrm{X1} < 10^{-7}$,
hence first-pass cyclotron damping is negligible in the underdense edge of CNT. 

\begin{figure} \begin{center}
\begin{center}
    \includegraphics[width=0.47\textwidth]{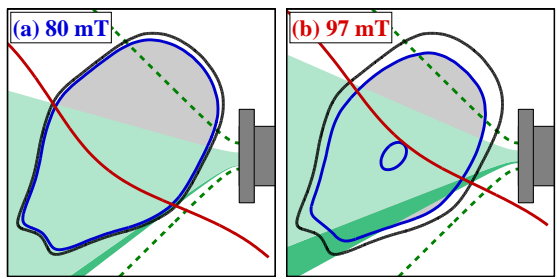}
    \caption{First-pass access to fundamental EC resonance 
             for O-mode launch at different magnetic field
             strengths corresponding to 
             Fig.~\ref{fig:o1_modb_scan}d-e. A large 
             fraction of the beam (light green) has to travel through the
             evanescent overdense region (gray) inside the
             O-mode cutoff (blue) before
             reaching the resonance (red). The darker green stripe is 
             the portion of the beam which reaches the EC resonance before 
             encountering the O-mode cutoff.}
    \label{fig:modBscan_schematic}
\end{center}
\end{center} \end{figure}

On the other hand, some cyclotron damping 
may occur in the overdense core.
This is attractive because it intercepts a larger and more
central (thus, more intense) portion of the Gaussian beam.
The interceptions are 
schematically illustrated in Fig.~\ref{fig:modBscan_schematic}
in the two limits of low and high field
(Figs.~\ref{fig:x1_modb_scan}d-e), corresponding to different
locations of the EC resonance. 
Cyclotron damping of the attenuated O- or X-mode 
is possible, in the overdense core, provided that the wave 
reaches the Doppler-broadened EC resonance with finite wave-amplitude and,
again, provided that the optical depth is sufficient. 
The increased value of $\omega_{pe}/\omega_{ce}$ is advantageous in this
respect. However, it is not as high as
to compensate for the five orders of magnitude mentioned before.
Moreover, the wave evanesces too rapidly. 
As an example, let us consider the 0.5 kW O-mode case
in Fig.~\ref{fig:o1_pwr_scan}, where $n_e \simeq 1\times10^{17}~\mathrm{m}^{-3}$,
and let us consider a location where $|\mathbf{B}| \simeq 90$ mT and  
$\theta \simeq 90^\circ$.  
The Appleton-Hartree dispersion relation\cite{stix1992} gives us a  
wavenumber $k \approx 30i$ for O-mode propagation;
hence, the electric field decays 
to $1/e$ of its incident value within 3 cm of the cutoff layer. The penetration
depth would be even lower in denser plasmas. 

This is in contradiction with O-mode experiments, where denser plasmas
tend to be hotter (Figs.~\ref{fig:o1_pwr_scan}a-b and
\ref{fig:o1_modb_scan}a-b). 
Due to this contradiction and, more importantly, to the low 
optical thickness and short penetration length, it is concluded that cyclotron 
damping of the O- and X-mode is not a plausible heating mechanism, neither
in the underdense nor in the overdense regions. 



\subsection{Collisional absorption at the UHR vs.~excitation of EBWs}
\label{subsect:collAbs}

In Secs.~\ref{subsect:SX-B}-\ref{subsect:O-X-B} we have discussed three
mechanisms by which the X-mode can reach the UHR despite cutoffs. 
For brevity they were referred to by their standard names in the literature:
SX-B (unlikely), FX-B (most likely) and O-X-B (somewhat likely) mode
conversions. 
It remains to be discussed whether the {\em electrostatic} B mode actually
develops. This mode has wavelength comparable with the electron Larmor
radius $r_\mathrm{Le}$, 
propagates {\em away} from the UHR (in this heating scheme; the opposite
in a diagnotic scheme) and is {\em cyclotron}-damped near the EC
resonance.

The alternative is that an {\em electromagnetic} SX mode 
{\em approaches} the UHR, decelerates, acquires shorter and
shorter wavelength but does not convert in EBWs, because first it is
{\em collisionaly} damped near the UHR. 
Note that even in the so called FX-B conversion, the FX branch couples 
with the SX branch of the dispersion relation first, before converting in EBWs.

In CNT, like in other cold plasmas or plasma edges,
collisional damping prevails on the conversion to EBWs.
This is because the electron-ion collisionality is relatively high
($\nu_\mathrm{ei} = $ 1-10 MHz): collisions are numerous, during the long 
time spent by the {\em slow} X wave near the UHR.  
The time spent near the UHR
can be estimated numerically, but, as an order of magnitude, the
Larmor radius for $T_e =$ 5 eV and $|\mathbf{B}|$ = 70 mT is
$r_\mathrm{Le} = 80~\mu\mathrm{m}$.
Waves of frequency $f$ = 2.45 GHz 
and wavelength $\lambda = 5 r_\mathrm{Le}$ propagate with phase velocity
$v_p = 10^{6}$ m/s. By the time this wave has
travelled 10 cm near the UHR, on average every electron has collided with an
ion once, in a plasma of $\nu_\mathrm{ei} = $ 10 MHz.
Here ``every electron'' includes electrons 
oscillating in the wave field (upper hybrid oscillations) and,
effectively, sustaining these partly electrostatic waves. Due to collisions,
energy is transfered from these electrons (thus, ultimately, from the wave)
to the {\em ions}, which are not resonating with the
{\em electron} Bernstein wave and are not supporting it.
In brief, electron-ion collisions subtract energy from the wave. 
Here this effect is quantified with the aid of the IPF-FDMC full-wave
code that, as mentioned, includes collisions.

\subsection{Earlier experiments in other devices}

Similar mechanisms have been observed on other devices with 
similar plasma parameters. Full-wave simulations and modulated power 
measurements
for 2.45 GHz plasmas in TJ-K both indicated that power deposition occurred
primarily at the UHR for both O- and X-mode 
heating.\cite{koehn2010} In WEGA, simulations and experiments both pointed to
significant deposition of EBWs at the cyclotron resonance.\cite{podoba2007}
The TJ-K experimental conditions are perhaps more similar
to CNT because of the steep density gradients in the vicinity of 
the upper hybrid resonance. The WEGA setup, by contrast, (1) was designed 
to place the UHR in a region with a low $n_e$ gradient and
(2) made use of an antenna optimized for O-X coupling. Power modulation 
measurements on CHS indicated significant levels of EBW heating
for both O- and X-mode (linearly polarized), injected both normally and 
obliquely to the O-cutoff.\cite{ikeda2008}

\section{Conclusions and future work}

In summary, overdense microwave plasma heating has been observed in CNT
with $n_e$ exceeding 
the cutoff density by factors of more than 4.
Density and temperature profiles
tend to be hollow, and changes in heating power affect temperatures primarily
at the edge rather than at the core. Variations in $|\mathbf{B}|$ had 
significant effects on the profiles for O-mode launch, but not for X-mode
launch.

These observations are consistent with collisional damping of an X-mode at
the upper hybrid resonance (UHR). Note that the
X-mode can reach the UHR even for O-mode launch, after O-X mode conversion
and/or polarization-scrambling reflections off the CNT walls and
internal coils. Such interpretation is supported by full-wave modeling
performed with the IPF-FDMC code. 
In the relatively cold plasmas presented, heated with less than 10 kW of
microwave power, the X-mode is completely damped before
completing its conversion in the Bernstein mode.

Future work will include upgrades to higher power for high $\beta$
stability research \cite{hammond_pop2017}, as well as
modulated-power experiments that will improve
our understanding of the heating deposition locations. 
In addition, full-wave calculations in three dimensions (3D) 
will further refine the 2D modeling
presented here. 3D effects are expected to be
important, due to the low aspect ratio of CNT. 

\section{Acknowledgments}

The authors thank 
S.D.~Massidda and X.~Sarasola for their help in installing
an earlier 1 kW magnetron and obtaining the first microwave plasma,
which encouraged the experiments at $\le$ 10 kW presented here.

\bibliographystyle{unsrt} 

\begin{thebibliography}{10}

\bibitem{prater2004}
R.~Prater.
\newblock Heating and current drive by electron cyclotron waves.
\newblock {\em Phys. Plasmas}, 11(5):2349, 2004.

\bibitem{laqua2007}
H.~P. Laqua.
\newblock Electron bernstein wave heating and diagnostic.
\newblock {\em Plasma Physics and Controlled Fusion}, 49:R1, 2007.

\bibitem{maekawa2001}
T.~Maekawa, T.~Kobayashi, S.~Yamaguchi, K.~Yoshinaga, H.~Igami, M.~Uchida,
  H.~Tanaka, M.~Asakawa, and Y.~Terumichi.
\newblock Doppler-shifted cyclotron absorption of electron {Bernstein} waves
  via {$N_\parallel$}-upshift in a tokamak plasma.
\newblock {\em Physical Review Letters}, 86:3783, 2001.

\bibitem{shevchenko2002}
V.~Shevchenko, Y.~Baranov, M.~O'Brien, and A.~Saveliev.
\newblock {Generation of noninductive current by electron Bernstein waves on
  the COMPASS-D tokamak}.
\newblock {\em Physical Review Letters}, 89:265005, 2002.

\bibitem{yoshimura2013}
Y.~Yoshimura, H.~Igami, S.~Kubo, T.~Shimozuma, H.~Takahashi, M.~Nishiura,
  S.~Ohdachi, K.~Tanaka, K.~Ida, M.~Yoshinuma, C.~Suzuki, S.~Ogasawara,
  R.~Makino, H.~Idei, R.~Kumazawa, T.~Mutoh, H.~Yamada, and {the LHD Experiment
  Group}.
\newblock Electron bernstein wave heating by electron cyclotron wave injection
  from the high-field side in {LHD}.
\newblock {\em Nuclear Fusion}, 53:063004, 2013.

\bibitem{ikeda2008}
R.~Ikeda, K.~Toi, M.~Takeuchi, C.~Suzuki, T.~Shoji, T.~Akiyama, M.~Isobe,
  S.~Hishimura, S.~Okamura, K.~Matsuoka, and {CHS Experimental Group}.
\newblock Production and heating of overdense plasmas by mode-converted
  electron {Bernstein} waves at very low toroidal field in the {Compact}
  {Helical} {System}.
\newblock {\em Physics of Plasmas}, 15:072505, 2008.

\bibitem{seltzman2016}
A.~H. Seltzman, J.~K. Anderson, A.~M. DuBois, A.~Almagri, and C.~B. Forest.
\newblock X-ray analysis of electron bernstein wave heating in mst.
\newblock {\em Review of Scientific Instruments}, 87:11E329, 2016.

\bibitem{preinhaelter1973}
J.~Preinhaelter and V.~Kopeck{\'y}.
\newblock Penetration of high-frequency waves into a weakly inhomogeneous
  magnetized plasma at oblique incidence and their transformation to
  {Bernstein} modes.
\newblock {\em Journal of Plasma Physics}, 10:1, 1973.

\bibitem{laqua2003}
H.~P. Laqua, H.~Maassberg, F.~Volpe, and {the W7-AS Team and ECRH-Group}.
\newblock Electron bernstein wave heating and current drive in overdense
  plasmas in the {W7-AS} stellarator.
\newblock {\em Nuclear Fusion}, 43:1324, 2003.

\bibitem{mueck2007}
A.~{Mueck}, L.~{Curchod}, Y.~{Camenen}, S.~{Coda}, T.~P. {Goodman}, H.~P.
  {Laqua}, A.~{Pochelon}, L.~{Porte}, and F.~{Volpe}.
\newblock {Demonstration of Electron-Bernstein-Wave Heating in a Tokamak via
  O-X-B Double-Mode Conversion}.
\newblock {\em Phys. Rev. Lett.}, 98(17):175004, April 2007.

\bibitem{nagasaki2016}
K.~Nagasaki, Y.~Nakamura, S.~Kamioka, H.~Igami, F.~Volpe, T.~Stange,
  K.~Sakamoto, H.~Okada, T.~Minami, S.~Kado, S.~Kobayashi, S.~Yamamoto,
  S.~Ohshima, G.~Weir, S.~Konoshima, N.~Kenmochi, Y.~Otan, Y.~Yoshimura,
  N.~Marushchenko, and T.~Mizuuchi.
\newblock Development of electron bernstein emission diagnostic for heliotron
  j.
\newblock {\em Plasma Fusion Research}, 11:2402095, 2016.

\bibitem{koehn2010}
A.~K{\"o}hn, G.~Birkenmeier, E.~Holzhauer, M.~Ramisch, and U.~Stroth.
\newblock Generation and heating of toroidally confined overdense plasmas with
  2.45 {GHz} microwaves.
\newblock {\em Plasma Physics and Controlled Fusion}, 52:035003, 2010.

\bibitem{podoba2007}
Y.~Y. Podoba, H.~P. Laqua, G.~B. Warr, M.~Schubert, M.~Otte, S.~Marsen,
  F.~Wagner, and E.~Holzhauer.
\newblock Direct observation of electron-{Bernstein} wave heating by
  {O-X-B}-mode conversion at low magnetic field in the {WEGA} stellarator.
\newblock {\em Physical Review Letters}, 98:255003, 2007.

\bibitem{pedersen2004}
T.~S. Pedersen, A.~H. Boozer, J.~P. Kremer, R.~G. Lefrancois, W.~T. Reiersen,
  F.~D. Dahlgren, and N.~Pomphrey.
\newblock The {Columbia} {Nonneutral} {Torus}: a new experiment to confine
  nonneutral and positron-electron plasmas in a stellarator.
\newblock {\em Fusion Sci. Technol.}, 46:200, 2004.

\bibitem{kremer2006}
J.~P. Kremer, T.~S. Pedersen, R.~G. Lefrancois, and Q.~Marksteiner.
\newblock Experimental confirmation of stable, small-debye-length,
  pure-electron-plasma equilibria in a stellarator.
\newblock {\em Phys. Rev. Lett.}, 97:095003, 2006.

\bibitem{sarasola2012}
X.~Sarasola and T.~S. Pedersen.
\newblock First experimental studies of the physics of plasmas of arbitrary
  degree of neutrality.
\newblock {\em Plasma Phys. Controlled Fusion}, 54:124008, 2012.

\bibitem{hammond2016}
K.~C. Hammond, A.~Anichowski, P.~W. Brenner, T.~S. Pedersen, S.~Raftopoulos,
  P.~Traverso, and F.~A. Volpe.
\newblock Experimental and numerical study of error fields in the {CNT}
  stellarator.
\newblock {\em Plasma Physics and Controlled Fusion}, 58:074002, 2016.

\bibitem{hammond_pop2017}
K.~C. Hammond, S.~A. Lazerson, and F.~A. Volpe.
\newblock High-$\beta$ equilibrium and ballooning stability of the low aspect
  ratio {CNT} stellarator.
\newblock {\em Physics of Plasmas}, 24:042510, 2017.

\bibitem{hammond_rsi2016}
K.~C. Hammond, R.~R. Diaz-Pacheco, Y.~Kornbluth, F.~A. Volpe, and Y.~Wei.
\newblock Onion-peeling inversion of stellarator images.
\newblock {\em Review of Scientific Instruments}, 87:11E119, 2016.

\bibitem{brenner2008}
P.~W. Brenner, T.~S. Pedersen, J.~W. Berkery, Q.~R. Marksteiner, and M.~S.
  Hahn.
\newblock Magnetic surface visualizations in the {Columbia} {Non-Neutral}
  {Torus}.
\newblock {\em IEEE Transactions on Plasma Science}, 36(4):1108, 2008.

\bibitem{jaenicke1993}
R.~Jaenicke, E.~Ascasibar, P.~Grigull, I.~Lakicevic, A.~Weller, M.~Zippe,
  H.~Hailer, and K.~Schw{\"o}rer.
\newblock Detailed investigation of the vacuum magnetic surfaces on the {W7-AS}
  stellarator.
\newblock {\em Nuclear Fusion}, 33(5):687--704, 1993.

\bibitem{lazerson2016}
S.~Lazerson, M.~Otte, S.~Bozhnekov, C.~Biedermann, T.~S. Pedersen, and the
  W7-X~team.
\newblock First measurements of error fields on {W7-X} using flux surface
  mapping.
\newblock {\em Nuclear Fusion}, 56:106005, 2016.

\bibitem{stangeby2000}
P.~C. Stangeby.
\newblock {\em The Plasma Boundary of Magnetic Fusion Devices}.
\newblock Plasma Physics Series. Taylor \& Francis Group, 2000.

\bibitem{koehn2008}
A.~K{\"o}hn, {\'A}.~Cappa, E.~Holzhauer, F.~Castej{\'o}n, {\'A}.~Fern{\'a}ndez,
  and U.~Stroth.
\newblock Full-wave calculation of the {O-X-B} mode conversion of {Gaussian}
  beams in a cylindrical plasma.
\newblock {\em Plasma Physics and Controlled Fusion}, 50:085018, 2008.

\bibitem{tanga1986}
A.~Tanga, P.~R. Thomas, J.~G. Cordey, J.~P. Christiansen, S.~Ejima, A.~Kellman,
  E.~Lazzaro, P.~J. Lomas, P.~Morgan, M.~F. Nave, P.~Noll, and F.~C.
  Sch{\"u}ller.
\newblock Start-up of the {Ohmic} phase in {JET}.
\newblock In H.~Knoepfel, editor, {\em Tokamak Start-up}, page 159. Plenum
  Press, 1986.

\bibitem{yamada2005}
H.~{Yamada}, J.~H. {Harris}, A.~{Dinklage}, E.~{Ascasibar}, F.~{Sano},
  S.~{Okamura}, J.~{Talmadge}, U.~{Stroth}, A.~{Kus}, S.~{Murakami},
  M.~{Yokoyama}, C.~D. {Beidler}, V.~{Tribaldos}, K.~Y. {Watanabe}, and
  Y.~{Suzuki}.
\newblock {Characterization of energy confinement in net-current free plasmas
  using the extended International Stellarator Database}.
\newblock {\em Nucl.~Fusion}, 45:1684, 2005.

\bibitem{dressel2011}
M.~Dressel, O.~Klein, Donovan S., and G.~Gr\"uner.
\newblock High frequency resonant techniques for studying the complex
  electrodynamics response in solids.
\newblock {\em Ferroelectrics}, 176(1):285, 1996.

\bibitem{albajar2005}
F.~Albajar, M.~Bornatici, and E.~Engelmann.
\newblock Electron cyclotron radiative transfer in the presence of polarization
  scrambling in wall reflections.
\newblock {\em Nuclear Fusion}, 45:L9, 2005.

\bibitem{BeckSpizz}
P.~Beckmann and A.~Spizzichino.
\newblock {\em The Scattering of Electromagnetic Waves from Rough Surfaces}.
\newblock Radar Library. Artech House, 1987.

\bibitem{liu2015}
L.~Liu, X.~Li, and K.~Nonaka.
\newblock {\em Rev.\ Sci.\ Instrum.}, 86:023107, 2015.

\bibitem{budden1961}
K.~G. Budden.
\newblock {\em Radio Waves in the Ionosphere}.
\newblock Cambridge University Press, 1961.

\bibitem{stix1992}
T.~H. Stix.
\newblock {\em Waves in Plasmas}.
\newblock Springer, 1992.

\bibitem{mjohlus1984}
E.~Mj{\o}hlus.
\newblock Coupling to {Z} mode near critical angle.
\newblock {\em Journal of Plasma Physics}, 31:7, 1984.

\bibitem{igami2006}
H.~Igami, H.~Tanaka, and T.~Maekawa.
\newblock A survey of mode-conversion transparency windows between external
  electromagnetic waves and electron bernstein waves for various plasma slab
  boundaries.
\newblock {\em Plasma Physics and Controlled Fusion}, 48:573, 2006.

\bibitem{bornatici1983}
M.~Bornatici, R.~Cano, O.~De Barbieri, and F.~Engelman.
\newblock Electron cyclotron absorption and emission in fusion plasmas.
\newblock {\em Nucl. Fusion}, 23(9):1153--1257, 1983.

\end{thebibliography}

\end{document}